\documentclass{elsart}

\usepackage[square,comma]{natbib}
\usepackage{graphicx}
\usepackage{pxfonts}

\usepackage{amssymb}

\journal{}

\begin{document}

\thispagestyle{empty}
\begin{Large}
\textbf{DEUTSCHES ELEKTRONEN-SYNCHROTRON}

\textbf{\large{in der HELMHOLTZ-GEMEINSCHAFT}\\}
\end{Large}

DESY 07-058

May 2007

\begin{eqnarray}
\nonumber &&\cr \nonumber && \cr \nonumber &&\cr
\end{eqnarray}
\begin{eqnarray}
\nonumber
\end{eqnarray}
\begin{center}
\begin{Large}
\textbf{Theory of Nonlinear Harmonic Generation in Free-Electron
Lasers with Helical Wigglers}
\end{Large}
\begin{eqnarray}
\nonumber &&\cr \nonumber && \cr
\end{eqnarray}

\begin{large}
Gianluca Geloni, Evgeni Saldin, Evgeni Schneidmiller and Mikhail
Yurkov
\end{large}
\textsl{\\Deutsches Elektronen-Synchrotron DESY, Hamburg}
\begin{eqnarray}
\nonumber
\end{eqnarray}
\begin{eqnarray}
\nonumber
\end{eqnarray}
\begin{eqnarray}
\nonumber
\end{eqnarray}
ISSN 0418-9833
\begin{eqnarray}
\nonumber
\end{eqnarray}
\begin{large}
\textbf{NOTKESTRASSE 85 - 22607 HAMBURG}
\end{large}
\end{center}
\clearpage
\newpage

\begin{frontmatter}



\title{Theory of Nonlinear Harmonic Generation in Free-Electron Lasers with Helical Wigglers}


\author{Gianluca Geloni,}
\author{Evgeni Saldin,}
\author{Evgeni Schneidmiller}
\author{and Mikhail Yurkov}

\address{Deutsches Elektronen-Synchrotron (DESY), Hamburg,
Germany}

\begin{abstract}
Coherent Harmonic Generation (CHG), and in particular Nonlinear
Harmonic Generation (NHG), is of importance for both short
wavelength Free-Electron Lasers (FELs), in relation with the
achievement of shorter wavelengths with a fixed electron-beam
energy, and high-average power FEL resonators, in relation with
destructive effects of higher harmonics radiation on mirrors.  In
this paper we present a treatment of NHG from helical wigglers
with particular emphasis on the second harmonic. Our study is
based on an exact analytical solution of Maxwell's equations,
derived with the help of a Green's function method. In particular,
we demonstrate that nonlinear harmonic generation (NHG) from
helical wigglers vanishes on axis. Our conclusion is in open
contrast with results in literature, that include a kinematical
mistake in the description of the electron motion.
\end{abstract}


\end{frontmatter}


\clearpage

\section{\label{sec:intro} Introduction}

The study of Coherent Harmonic Generation (CHG) is of undisputed
relevance in the field of Free-Electron Lasers (FELs). On the one
hand, x-ray light sources based on Self-Amplified Spontaneous
Emission (SASE)  can benefit from CHG to radiate at shorter
wavelengths at the same electron beam energy (see e.g.
\cite{XFEL}). On the other hand, CHG can be a detrimental effect
for high average power FEL oscillators, because of possible mirror
damage from harmonics in the ultraviolet \cite{FRE2}. Thus,
correct understanding of CHG is of interest for both short
wavelengths and high average power applications, that are the two
major up-to-date development paths in FEL physics.

CHG is driven by bunching of the electron beam at harmonics of the
fundamental, and is characterized by the fact that harmonic
components of the bunched beam radiate coherently \cite{HUAN}. The
bunching mechanism may be linear or nonlinear. In the first case
CHG is named Linear Harmonic Generation (LHG). LHG arises when
radiation at a certain harmonic induces electron beam bunching at
the same harmonic. In the second case CHG is named Nonlinear
Harmonic Generation (NHG).  NHG arises when the intensity of the
fundamental harmonic\footnote{Note that this argument is not
restricted to the first harmonic. Here we have in mind an FEL
lazing at the fundamental.} is strong enough to induce bunching of
electrons at different (higher) harmonics. In FEL processes,
harmonic bunching of the electron beam due to interactions with
the fundamental is always much stronger than that due to
interactions with higher radiation harmonics. As a result only NHG
has practical relevance, while LHG can be neglected.  In general,
CHG can be treated in terms of an electrodynamical problem where
Maxwell's equations are solved with given sources in the
space-frequency domain. Sources must still be obtained through the
solution of self-consistent equations for electrons and fields.
However, once these equations are solved one obtains macroscopic
current and charge density distributions as a function of
transverse and longitudinal coordinates at a given harmonic.
Further on, solution of Maxwell's equations with these
distributions as given sources characterizes harmonic radiation in
the space-frequency domain. The dependence of sources in the
space-frequency domain on transverse and longitudinal coordinates
is complicated because is the result of the above-mentioned
self-consistent process. However, here we deal with an FEL setup
where an ultrarelativistic electron beam is sent, in free space,
through an undulator with many periods. Then, paraxial and
resonance approximation can be applied to simplify the
characterization of CHG. In particular, for a fixed transverse
position, the longitudinal dependence is always slow on the scale
of an undulator period.

NHG has been dealt with in the case of a planar wiggler, both
theoretically and experimentally, in a number of works
\cite{FRE2}-\cite{HAR2}. Odd harmonics have maximal power on
axis\footnote{Here we assume that the bunching wavefront is
perpendicular to the (longitudinal) FEL axis.} and are linearly
polarized. Even harmonics have been shown to have vanishing
on-axis power and to exhibit both horizontal and vertical
polarization components.

In this work we present the first exact theory of NHG from helical
wigglers. Our treatment is based on an exact solution of Maxwell's
equations in the space-frequency domain based on a Green's
function technique. In the next Section \ref{sec:theor}
electromagnetic sources are treated in all generality as given
data to be obtained from self-consistent FEL codes. In particular,
we will demonstrate that nonlinear harmonic generation (NHG) from
helical wigglers vanishes on axis. Later on we will focus on the
second harmonic, and discuss a particular study-case. Our results
are in contrast with conclusions in \cite{FREU}, where NHG in a
helical wiggler has also been addressed and the presence of
on-axis power has been reported. In Section \ref{sec:criti} we
will show that the azimuthal resonance condition introduced in
\cite{FREU} to explain net energy exchange between particles and
fields is a misconception. It arises from a kinematical mistake.
Namely, the electron rotation angle in the helical wiggler is
confused with the azimuthal coordinate of the cylindrical
reference system. Consequently, this misconception is passed on to
simulations, leading to incorrect results. We conclude our work
with some final remarks in Section \ref{sec:conc}.

\section{\label{sec:theor} Theory of nonlinear  harmonic
generation in helical wigglers}

\subsection{\label{sub:gen} Complete analysis of the harmonic generation mechanism}

In this Section we propose the first exact theory of NHG in
helical wigglers. In particular we will give an analysis in the
space-frequency domain of the second harmonic generation case.

Paraxial Maxwell's equations in the space-frequency domain can be
used to describe radiation from ultra-relativistic electrons (see
\cite{OURF,OURI}). Let us call the transverse electric field in
the space-frequency domain $\vec{\bar{E}}_\bot
(z,\vec{r}_\bot,\omega)$, where $\vec{r}_\bot = x
\vec{e}_x+y\vec{e}_y$ identifies a point on a transverse plane at
longitudinal position $z$, $\vec{e}_x$ and $\vec{e}_y$ being unit
vectors in the transverse $x$ and $y$ directions. Here the
frequency $\omega$ is related to the wavelength $\lambda$ by
$\omega = 2 \pi c/\lambda$, $c$ being the speed of light in
vacuum. From the paraxial approximation follows that the electric
field envelope $\vec{\widetilde{E}}_\bot = \vec{\bar{E}}_\bot
\exp{[-i\omega z/c]}$ does not vary much along $z$ on the scale of
the reduced wavelength $\lambdabar=\lambda/(2\pi)$. As a result,
the following field equation holds:

\begin{eqnarray} \mathcal{D}
\left[\vec{\widetilde{E}}_\bot(z,\vec{r}_\bot,\omega)\right] =
\vec{f}(z, \vec{r}_\bot,\omega) ~,\label{field1}
\end{eqnarray}
where the differential operator $\mathcal{D}$ is defined by

\begin{eqnarray}
\mathcal{D} \equiv \left({\nabla_\bot}^2 + {2 i \omega \over{c}}
{\partial\over{\partial z}}\right) ~,\label{Oop}
\end{eqnarray}
${\nabla_\bot}^2$ being the Laplacian operator over transverse
cartesian coordinates. Eq. (\ref{field1}) is Maxwell's equation in
paraxial approximation. The source-term vector $\vec{f}(z,
\vec{r}_\bot)$ is specified by the trajectory of the source
electrons, and can be written in terms of the Fourier transform of
the transverse current density, $\vec{\bar{j}}_\bot
(z,\vec{r}_\bot,\omega)$, and of the charge density,
$\bar{\rho}(z,\vec{r}_\bot,\omega)$, as

\begin{eqnarray}
\vec{f} = && - {4 \pi} \left(\frac{i\omega}{c^2}\vec{\bar{j}}_\bot
-\vec{\nabla}_\bot \bar{\rho}\right) \exp\left[-\frac{i \omega
z}{c}\right] ~. \label{fv}
\end{eqnarray}
In this paper we will treat $\vec{\bar{j}}_\bot$ and $\bar{\rho}$
as macroscopic quantities, without investigating individual
electron contributions. $\vec{\bar{j}}_\bot$ and $\bar{\rho}$ are
regarded as given data, that can be obtained from any FEL code.
Codes actually provide the charge density of the modulated
electron beam in the time domain $\rho(z,\vec{r}_\bot,t)$. A
post-processor can then be used in order to perform the Fourier
transform of $\rho$, that can always be presented as

\begin{eqnarray}
\bar{\rho} = - \widetilde{\rho}(z,\vec{r}_\bot-\vec{r'}_{o
\bot}(z),\omega) \exp\left[
i\omega\frac{s_o(z)}{v_o}\right]~,\label{rhotr}
\end{eqnarray}
where the minus sign on the right hand side is introduced for
notational convenience only. Quantities $\vec{r'}_{o \bot}(z)$,
$s_o(z)$ and $v_o$ pertain a reference electron with nominal
Lorentz factor $\gamma_o$ that is injected on axis with no
deflection and is guided by the helical undulator field. Such
electron follows a helical trajectory $\vec{r'}_{o\bot}(z)=
r'_{ox} \vec{e}_x+r'_{oy} \vec{e}_y$. We assume that the reference
electron rotates anti-clockwise in the judgement of an observer
located after the undulator and looking towards the device, so
that

\begin{eqnarray}
&&r'_{ox}(z) = \frac{ K }{\gamma_o k_w} [\cos(k_w z)-1]  ~~,~~~
r'_{oy}(z) = \frac{K}{\gamma_o k_w} \sin(k_w z) ~. \label{rhel0}
\end{eqnarray}
In Eq. (\ref{rhel0}), $K=\lambda_w e H_w/(2\pi m_e c^2)$ is the
undulator parameter, $\lambda_w = 2\pi/k_w$ being the undulator
period, $(-e)$ the negative electron charge, $H_w$ the maximal
modulus of the undulator magnetic field on-axis, and $m_e$ the
rest mass of the electron.  The corresponding velocity is
described by $\vec{v}_{o\bot}(z)=  v_{ox} \vec{e}_x+v_{oy}
\vec{e}_y$ with

\begin{eqnarray}
&&v_{ox}(z) = -  \frac{ K c}{\gamma_o} \sin(k_w z) ~~,~~~v_{oy}(z)
= ~~\frac{K c}{\gamma_o} \cos(k_w z) ~. \label{vhel0}
\end{eqnarray}
Finally, $s_o(z)$ is the curvilinear abscissa measured along the
trajectory of the reference particle.

One may always present $\bar{\rho}$ as in Eq. (\ref{rhotr}).
However, introduction of $\widetilde{\rho}$ is useful when
$\widetilde{\rho}$ is a slowly varying function of $z$ on the
wavelength scale. This property is granted by the fact that the
charge density distribution under study originates from an FEL
process. From this fact it also follows that $\widetilde{\rho}$ is
slowly varying on the scale of the undulator period $\lambda_w$
and is peaked around each harmonic of the fundamental $\omega_r =
2 k_w c \bar{\gamma}_z^2$, that is fixed imposing resonance
condition between electric field and reference particle. The word
"peaked" means that the bandwidth of each harmonic component obeys
$\Delta \omega/(h \omega_r) \ll 1$ for each positive integer value
$h$. Here $\bar{\gamma}_z  = 1/(1-v_{oz}^2/c^2)$ is the
longitudinal Lorentz factor. Note that, for the reference
electron, $\bar{\gamma}_z$ does not depend on $z$. We have

\begin{equation}
\bar{\gamma}_z \equiv \frac{1}{\sqrt{1-v_{oz}^2/c^2}} =
\frac{\gamma}{\sqrt{1+K^2}}~, \label{gamzdef}
\end{equation}
where the last equality follows from $v_{oz}^2 = v_o^2 -
v_{o\bot}^2$, together with Eq. (\ref{vhel0}). Finally, the
relative deviation of the particles energy from $\gamma_o m_e c^2$
is small, i.e. $\delta \gamma/\gamma_o \ll 1$.

In this paper we will be interested in the case when the
transverse beam dimension $\sigma_\bot$ is much larger than the
electron rotation radius $r_w$, i.e. $\sigma_\bot \gg r_w$. We
might then substitute the dependence on $\vec{r}_\bot-\vec{r'}_{o
\bot}(z)$ in Eq. (\ref{rhotr}) with a dependence on
$\vec{r}_\bot$, because the electron rotation radius is negligible
with respect to $\sigma_\bot$ and individual electrons can be
considered as occupying a fixed transverse position. However, we
will not do so. In fact, starting from Eq. (\ref{rhotr}) to
develop our theory, we will be able to crosscheck our results for
an extended source with the well-known asymptotic for a filament
beam (see Section \ref{sub:mod}), i.e. $\sigma_\bot \ll r_w$, and
demonstrate agreement with results in \cite{KIN1}.

We note that for a generic motion one has

\begin{equation}
\omega \left({s(z_2)-s(z_1)\over{v}}-{z_2-z_1\over{c}}\right) =
\int_{z_1}^{z_2} d \bar{z} \frac{\omega}{2 \gamma_z^2(\bar{z})
c}~, \label{moregen}
\end{equation}
Thus, with the help of Eq. (\ref{rhotr}), Eq. (\ref{fv}) can be
presented as\footnote{Eq. (\ref{fvtf}) is the analogous of the
source term on the right hand side of Eq. (11) in reference
\cite{HAR2}, dealing with planar undulators and an electron beam
modulated at the second harmonic only. There is only a slight
notational difference in that the symbol $\widetilde{\rho}$ in the
present paper corresponds to $j_o \widetilde{a}_2/c$ in reference
\cite{HAR2}.}

\begin{eqnarray}
\vec{f} = && {4 \pi} \exp\left[i \int_{0}^{z} d \bar{z} \frac{
\omega }{2 \bar{\gamma}_z^2  c}\right]
\left[\frac{i\omega}{c^2}\vec{v}_{o\bot}(z) -\vec{\nabla}_\bot
\right]\widetilde{\rho}[z;\vec{r}_\bot - \vec{r'}_{o\bot}(z)]
~,\cr && \label{fvtf}
\end{eqnarray}
where we used the fact that $\vec{\bar{j}}_\bot = \vec{v}_{o\bot}
\bar{\rho}$. In fact, for each particle in the beam $\delta
\gamma/\gamma_o \ll 1$. Therefore we can neglect differences
between the average transverse velocity of electrons $\langle
\vec{v}_\bot \rangle$ and $\vec{v}_{o\bot}$, so that
$\vec{\bar{j}}_\bot \equiv \langle \vec{v}_\bot \rangle \bar{\rho}
\simeq \vec{v}_{o\bot} \bar{\rho}$.

We will now introduce a coherent deflection angle
$\vec{\eta}^{(c)}$, where the superscript $(c)$ stands for
"coherent", to describe the transverse deflection of the electron
beam as a whole\footnote{With this, we assume that the deflection
angle $\vec{\eta}^{(c)}$ is constant. This is the case only if we
do not account for focusing elements within the undulator.
Generalization to account for betatron motion of electrons is,
however, straightforward.}. This means that we account for a
possible deflection angle $\vec{\eta}^{(c)}$ in the the trajectory
of reference electron. We therefore perform the following
substitutions in Eq. (\ref{fvtf}):

\begin{eqnarray}
&&\vec{r'}_{o\bot}(z) \longrightarrow
\vec{r}_{\bot}^{(c)}(z,\vec{\eta}^{(c)}) = \cr
&&\vec{r'}_{o\bot}(z) + \vec{\eta}^{(c)} z =
\left\{\frac{K}{\gamma k_w }\left[\cos(k_w z)-1\right]
+\eta_x^{(c)} z \right\}\vec{e}_x + \left\{\frac{K}{\gamma k_w
}\sin(k_w z) +\eta_y^{(c)} z \right\}\vec{e}_y~,\cr &&
\label{etac}
\end{eqnarray}
\begin{eqnarray}
&&\vec{v}_{o\bot}(z) \longrightarrow
\vec{v}_{\bot}(z,\vec{\eta}^{(c)}) = \cr &&  \vec{v}_{o\bot}(z) +
c \vec{\eta}^{(c)} = \left[-\frac{Kc}{\gamma} \sin(k_w z) + c
\eta_x^{(c)}\right]\vec{e}_x+\left[\frac{Kc}{\gamma} \cos(k_w z) +
c \eta_y^{(c)}\right]\vec{e}_y~. \label{vetac}
\end{eqnarray}
Using $\vec{v}_{\bot}(z,\vec{\eta}^{(c)})$ in place of $
\vec{v}_{o\bot}(z)$ implies that $\gamma_z(z,\vec{\eta}^{(c)})$ is
now a function of both $z$ and $\vec{\eta}^{(c)}$. In particular,
$1/\gamma_z^2(z,\vec{\eta}^{(c)}) = 1 -
v_z^2(z,\vec{\eta}^{(c)})/c^2$, where $v_z^2 = v^2 - v_\bot^2$ is
the square of the electron longitudinal velocity. It follows that
$1/\bar{\gamma}_z^2$ in Eq. (\ref{fvtf}) should also be
substituted according to

\begin{eqnarray}
&& \frac{1}{\bar{\gamma}_z^2} \longrightarrow
\frac{1}{\gamma_z^2(z,\vec{\eta}^{(c)})} = \cr
&&\frac{1}{\bar{\gamma}_z^2} + \left[\left(\eta^{(c)}_x\right)^2+
\left(\eta^{(c)}_y\right)^2\right]  + \frac{2 K}{\gamma} \left[-
\eta^{(c)}_x \sin(k_w z) + \eta^{(c)}_y \cos(k_w z)\right]~, \cr
&& \label{gammac}
\end{eqnarray}
yielding

\begin{eqnarray}
\vec{f} = && {4 \pi} \exp\left[i \int_{0}^{z} d \bar{z}\frac{
\omega }{2c\gamma_z^2(\bar{z},\vec{\eta}^{(c)})} \right]
\left[\frac{i\omega}{c^2}\vec{v}_{\bot}(z,\vec{\eta}^{(c)})
-\vec{\nabla}_\bot \right]\widetilde{\rho}[z;\vec{r}_\bot -
\vec{r}^{(c)}_{\bot}(z,\vec{\eta}^{(c)})] ~.\cr && \label{fvtff}
\end{eqnarray}
We find an exact solution of Eq. (\ref{Oop}) without any other
assumption about the parameters of the problem. A Green's function
for Eq. (\ref{Oop}), namely the solution corresponding to the unit
point source can be written as (see \cite{HAR2}):

\begin{eqnarray}
G(z_o-z';\vec{r_{\bot o}}-\vec{r'_\bot}) &=& -{1\over{4\pi
(z_o-z')}} \exp\left\{ i\omega{\mid \vec{r_{\bot o}}
-\vec{r'_\bot}\mid^2\over{2c (z_o-z')}}\right\}\label{green}~,
\end{eqnarray}
assuming $z_o-z' > 0$. When $z_o-z' < 0$ the paraxial
approximation does not hold, and the paraxial wave equation Eq.
(\ref{field1}) should be substituted, in the space-frequency
domain, by a more general Helmholtz equation. However, the
radiation formation length for $z_o - z'<0$ is very short with
respect to the case $z_o - z' >0$, i.e. there is no radiation for
observer positions $z_o-z' <0$. As a result, in this paper we will
consider only $z_o - z'> 0$. It follows that the observer is
located downstream of the sources.

This leads to the solution

\begin{eqnarray}
&&\vec{\widetilde{E}}_{\bot }(z_o, \vec{r}_{\bot o} )=
-\int_{-\infty}^{\infty} dz' \frac{1}{z_o-z'} \int d
\vec{r'}_{\bot}
\left[\frac{i\omega}{c^2}\vec{v}_\bot(z',\vec{\eta^{(c)}})
-\vec{\nabla}'_\bot \right]\cr &&\times \widetilde{\rho}
\left(z',\vec{r'}_\bot -\vec{r}^{(c)}_\bot(z',\vec{\eta^{(c)}})
\right) \exp\left\{i\omega\left[\frac{\mid \vec{r}_{\bot
o}-\vec{r'}_\bot \mid^2}{2c (z_o-z')}\right]+ i\int_{0}^{z'} d
\bar{z}\frac{ \omega }{2
 c\gamma_z^2(\bar{z},\vec{\eta}^{(c)})} \right\} ~, \cr && \label{blob}
\end{eqnarray}
where $\vec{\nabla}'_\bot$ represents the gradient operator
(acting on transverse coordinates) with respect to the source
point, while $(z_o, \vec{r}_{\bot o})$ indicates the observation
point. Note that the integration is taken from $-\infty$ to
$\infty$ because the sources are understood to be localized in a
finite region of space (i.e. $\widetilde{\rho}$ is different from
zero in a finite region of space). This is in agreement with
$z_o-z'>0$ and within the applicability criteria of the paraxial
approximation. Integration by parts of the gradient term leads to

\begin{eqnarray}
\vec{\widetilde{E}}_{\bot}&= &-\frac{i \omega }{c}
\int_{-\infty}^{\infty} dz' \frac{1}{z_o-z'}  \int d
\vec{r'}_{\bot} \left(\frac{\vec{v}_\bot(z',\vec{\eta^{(c)}})}{c}
-\frac{\vec{r}_{\bot o}-\vec{r'}_\bot}{z_o-z'}\right)\cr&&\times
\widetilde{\rho} \left(z',\vec{r'}_\bot
-\vec{r}^{(c)}_\bot(z',\vec{\eta^{(c)}}) \right) \exp\left[i
\Phi_T(z',\vec{r'}_\bot,\vec{\eta^{(c)}})\right] ~, \cr &&
\label{generalfin}
\end{eqnarray}
where the total phase $\Phi_T$ is given by

\begin{equation}
\Phi_T =  \omega\left[\frac{\mid \vec{r}_{\bot o}-\vec{r'}_\bot
\mid^2}{2c (z_o-z')}\right]+  \int_{0}^{z} d \bar{z}\frac{ \omega
}{2 c \gamma_z^2(\bar{z},\vec{\eta}^{(c)})}~. \label{totph}
\end{equation}
We now make use of a new integration variable $\vec{l}'=
\vec{r'}_\bot-\vec{r}^{(c)}_\bot(z',\vec{\eta^{(c)}})$ so that

\begin{eqnarray}
\vec{\widetilde{E}}_{\bot }&=& -\frac{i \omega }{c}
\int_{-\infty}^{\infty} dz' \frac{1}{z_o-z'}  \int d \vec{l}'
\left(\frac{\vec{v}_\bot(z',\vec{\eta^{(c)}})}{c}
-\frac{\vec{r}_{\bot o}-
\vec{r}^{(c)}_\bot(z',\vec{\eta^{(c)}})-\vec{l}'}{z_o-z'}\right)\cr&&
\times  \widetilde{\rho} \left(z',\vec{l}' \right) \exp\left[i
\Phi_T(z',\vec{l}',\vec{\eta^{(c)}})\right] ~, \label{generalfin2}
\end{eqnarray}
and

\begin{equation}
\Phi_T =  \omega \left[ \frac{|\vec{r}_{\bot
o}-\vec{r}^{(c)}_\bot(z',\vec{\eta^{(c)}})-\vec{l}'|^2}{2c
(z_o-z')}\right]+ \int_{0}^{z'} d \bar{z}\frac{ \omega }{2
 c\gamma_z^2(\bar{z},\vec{\eta}^{(c)})}~. \label{totph2}
\end{equation}
We will be interested in the total power emitted and in the
directivity diagram of the radiation in the far zone.

We therefore introduce the far zone approximation calling the
observation angle $\vec{\theta}=\vec{r}_{\bot o}/z_o$, setting
$\theta \equiv |\vec{\theta}|$ and taking the limit for $z_o \gg
L_w$, where $L_w = N_w \lambda_w$ is the undulator length:

\begin{eqnarray}
&&\vec{\widetilde{E}}_{\bot }= -\frac{i \omega }{c z_o}
\int_{-\infty}^{\infty} dz'   \int d \vec{l}'~
\left(\frac{\vec{v}_\bot(z',\vec{\eta^{(c)}})}{c}
-\vec{\theta}\right)  \widetilde{\rho} \left(z',\vec{l}' \right)
\cr&& \times \exp\left\{\frac{i\omega}{2c} \left[z_o \theta^2 - 2
\vec{\theta}\cdot \vec{r}^{(c)}_\bot(z',\vec{\eta^{(c)}})- 2
\vec{\theta}\cdot \vec{l}' +z'\theta^2\right]+ i \int_{0}^{z'} d
\bar{z}\frac{ \omega }{2
 c\gamma_z^2(\bar{z},\vec{\eta}^{(c)})}\right\}~
.\cr && \label{generalfin3}
\end{eqnarray}
Substitution of Eq. (\ref{etac}), Eq. (\ref{vetac}) and Eq.
(\ref{gammac}) in Eq. (\ref{generalfin3}) yields the following
field contribution calculated along the undulator:

\begin{eqnarray}
\vec{\widetilde{E}}_{\bot} &=& \frac{i \omega }{c z_o} \int
d\vec{l}' \exp\left[i\Phi_o\right] \int_{-L_w/2}^{L_w/2} dz'
\widetilde{\rho} \left(z',\vec{l}' \right){\exp{\left[i
\Phi_T\right]}} \cr &&\times \left[\left(\frac{K}{\gamma}
\sin\left(k_w
z'\right)+\left(\theta_x-\eta^{(c)}_x\right)\right)\vec{e}_x
+\left(-\frac{K}{\gamma} \cos\left(k_w
z'\right)+\left(\theta_y-\eta^{(c)}_y\right)\right)\vec{e}_y\right]~,
\label{undurad}
\end{eqnarray}
where

\begin{eqnarray}
\Phi_T &=& \frac{\omega z'}{2 c}
\left[\frac{1}{\bar{\gamma}_z^2}+\left(\theta_x-\eta_x^{(c)}\right)^2+\left(\theta_y-\eta_y^{(c)}\right)^2\right]
\cr &&-\frac{K\omega}{c \gamma
k_w}\left[{(\theta_y-\eta^{(c)}_y)}\sin(k_w z')  +
{(\theta_x-\eta^{(c)}_x)}\cos(k_w z')\right] ~\label{phitundu}
\end{eqnarray}
and

\begin{eqnarray}
\Phi_o &=& \frac{\omega}{c} \left[
\frac{z_o\left(\theta_x^2+\theta_y^2\right)}{2} +\frac{K
(\theta_x-\eta_x^{(c)})}{k_w \gamma}- (\theta_x l'_x +\theta_y
l'_y)\right]~.\label{phitunduo}
\end{eqnarray}
Note that the integration in Eq. (\ref{undurad}) is performed in
$d {z}'$ over the undulator length, i.e. is limited to the
interval $[-L_w/2,L_w/2]$. The reason for this is that, working
under the resonance approximation in the limit for $N_w \gg 1$,
one can neglect contributions to the field due to non-resonant
elements outside the undulator \cite{OURI}.

We are interested in studying frequency near the fundamental
harmonic $\omega_r = 2 k_w c \bar{\gamma}_z^2$ or its $h$-th
integer multiple. We specify "how near" the frequency $\omega$ is
to the $h$-th harmonic by defining a detuning parameter $C_h$ as

\begin{eqnarray}
C_h = \frac{\omega}{2\bar{\gamma}_z^2 c} - h k_w = \frac{\Delta
\omega}{\omega_r} k_w~. \label{Ch}
\end{eqnarray}
Here $\omega = h \omega_r + \Delta \omega$. Eq. (\ref{phitundu})
can thus be rewritten as

\begin{eqnarray}
\Phi_T &=& z' \left[h k_w + C_h +\frac{\omega}{2 c}
\left(\theta_x-\eta_x^{(c)}\right)^2+\frac{\omega}{2
c}\left(\theta_y-\eta_y^{(c)}\right)^2\right] \cr
&&-\frac{K\omega}{c \gamma
k_w}\left[{(\theta_y-\eta^{(c)}_y)}\sin(k_w z')  +
{(\theta_x-\eta^{(c)}_x)}\cos(k_w z')\right] ~,\label{phitundu2}
\end{eqnarray}
so that altogether, Eq. (\ref{undurad}) can be presented as:

\begin{eqnarray}
\vec{\widetilde{E}}_{\bot} &=& \frac{i \omega }{c z_o}
\int_{-\infty}^{\infty} d l'_x \int_{-\infty}^{\infty} d l'_y
\int_{-L_w/2}^{L_w/2} dz' \widetilde{\rho} \left(z',l'_x,l'_y
\right) \cr &&\times \exp\left\{i \frac{\omega}{c} \left[
\frac{z_o\left(\theta_x^2+\theta_y^2\right)}{2} +\frac{K
\left(\theta_x-\eta_x^{(c)}\right)}{k_w \gamma}- \left(\theta_x
l'_x +\theta_y l'_y\right)\right] \right\} \cr && \times
\exp\left\{i \left[h k_w + C_h +\frac{\omega}{2 c}
\left(\theta_x-\eta_x^{(c)}\right)^2+\frac{\omega}{2
c}\left(\theta_y-\eta_y^{(c)}\right)^2\right] z' \right.\cr
&&\left.-i \frac{K\omega}{c \gamma
k_w}\left[{(\theta_y-\eta^{(c)}_y)}\sin(k_w z')  +
{(\theta_x-\eta^{(c)}_x)}\cos(k_w z')\right]\right\} \cr &&\times
\left\{\left[\frac{K}{2i \gamma} \left(\exp[i k_w z']-\exp[-i k_w
z'] \right)
+\left(\theta_x-\eta^{(c)}_x\right)\right]\vec{e}_x\right.\cr
&&\left. +\left[-\frac{K}{2\gamma}\left(\exp[i k_w z']+\exp[-i k_w
z']
\right)+\left(\theta_y-\eta^{(c)}_y\right)\right]\vec{e}_y\right\}~,
\label{unduradtot}
\end{eqnarray}
We make use of the well-known expansion (see \cite{ALFE})

\begin{equation}
\exp{[i a \sin{(\psi)}]}=\sum_{p=-\infty}^{\infty} J_p(a) \exp{[i
p \psi]}~, \label{alfeq} \end{equation}
where $J_p$ indicates the Bessel function of the first kind of
order $p$, while $a$ and $\psi$ are real numbers. Eq.
(\ref{unduradtot}) may thus be written as

\begin{eqnarray}
\vec{\widetilde{E}}_{\bot} &=& \frac{i \omega }{c z_o}
\int_{-\infty}^{\infty} d l'_x \int_{-\infty}^{\infty} d l'_y
\int_{-L_w/2}^{L_w/2} dz' \widetilde{\rho} \left(z',l'_x, l'_y
\right) \cr &&\times \exp\left\{i \frac{\omega}{c} \left[
\frac{z_o\left(\theta_x^2+\theta_y^2\right)}{2} +\frac{K
(\theta_x-\eta_x^{(c)})}{k_w \gamma}- (\theta_x l'_x +\theta_y
l'_y)\right] \right\} \cr && \times \sum_{m,n=-\infty}^{\infty}
J_m(u) J_n(v) \exp\left[\frac{i \pi n}{2}\right] \exp\left[i
(n+m+h) k_w z'\right] \cr && \times\exp\left\{i \left[C_h
+\frac{\omega}{2 c}
\left(\theta_x-\eta_x^{(c)}\right)^2+\frac{\omega}{2
c}\left(\theta_y-\eta_y^{(c)}\right)^2\right]z'\right\} \cr
&&\times \left\{\left[\frac{K}{2i \gamma} \left(\exp[i k_w
z']-\exp[-i k_w z'] \right)
+\left(\theta_x-\eta^{(c)}_x\right)\right]\vec{e}_x\right.\cr
&&\left. +\left[-\frac{K}{2\gamma}\left(\exp[i k_w z']+\exp[-i k_w
z']
\right)+\left(\theta_y-\eta^{(c)}_y\right)\right]\vec{e}_y\right\}~,
\label{undurads1}
\end{eqnarray}
where

\begin{equation}
u = -\frac{K  \omega (\theta_y-\eta_y^{(c)})}{ c \gamma k_w
}~~~~\mathrm{and}~~~v = -\frac{K  \omega (\theta_x-\eta_x^{(c)})}{
c \gamma k_w }~. \label{uv}
\end{equation}
Whenever

\begin{equation}
C_h +\frac{\omega}{2 c}\left[
\left(\theta_x-\eta_x^{(c)}\right)^2+\left(\theta_y-\eta_y^{(c)}\right)^2\right]
\ll k_w \label{eqq} ~,
\end{equation}
the second phase factor in $z'$ in Eq. (\ref{undurads1}) (the one
containing $C_h$) is varying slowly on the scale of the undulator
period $\lambda_w$. As a result, simplifications arise when $N_w
\gg 1$, because fast oscillating terms in powers of $\exp[i k_w
z']$ effectively average to zero.

Let us use  Eq. (\ref{undurads1}), to discuss harmonic radiation
characteristics. Within the resonance approximation we further
select frequencies such that

\begin{eqnarray}
\frac{|\Delta \omega|}{\omega_r} \ll 1~,~~~~ \mathrm{i.e.}~~|C_h|
\ll k_w ~.\label{resext}
\end{eqnarray}
Note that this condition on frequencies automatically selects
observation angles of interest $h
(\vec{\theta}-\vec{\eta}^{(c)})^2 \ll 1/\bar{\gamma}_z^2$. In
fact, if one considers observation angles outside this range,
condition (\ref{eqq}) is not fulfilled, and the integrand in Eq.
(\ref{undurads1}) exhibits fast oscillations on the integration
scale $L_w$. As a result, one obtains zero transverse field,
$\vec{\widetilde{E}}_\bot = 0$, with accuracy $1/N_w$. Under the
constraint imposed by (\ref{resext}), independently of the value
of $K$ and for observation angles of interest we have

\begin{equation}
|v|= \left(1+\frac{\Delta \omega}{h \omega_r}\right) \frac{2
K}{\sqrt{1+K^2}} \bar{\gamma}_z h |\theta_x-{\eta_x}^{(c)}|
\lesssim \bar{\gamma}_z h |\theta_x-{\eta_x}^{(c)}| \ll 1~
\label{dropv}
\end{equation}
and similarly

\begin{equation}
|u|= \left(1+\frac{\Delta \omega}{h \omega_r}\right) \frac{2
K}{\sqrt{1+K^2}} \bar{\gamma}_z h |\theta_y-{\eta_y}^{(c)}|
\lesssim \bar{\gamma}_z h |\theta_y-{\eta_y}^{(c)}|  \ll 1~.
\label{dropu}
\end{equation}
This means that, independently of $K$, $|v| \ll 1$ and $|u| \ll
1$, so that we may expand $J_n(v)$ and $J_m(u)$ in Eq.
(\ref{undurads1}) according to $J_p(x) \simeq S^p(p)
[2^{-|p|}/\Gamma(1+|p|)]~x^{|p|}$, where $S(p)$ gives the sign of
the integer $p$ if $p \ne 0$ and unity if $p=0$, while $\Gamma$ is
the Euler gamma function

\begin{eqnarray}
\Gamma(z) = \int_0^\infty dt~t^{z-1} \exp[-t] ~.\label{geule}
\end{eqnarray}
From now on we will restrict our attention to the second harmonic,
noting that we may study any harmonic of interest in a similar
way.

Thus, $h=2$. Terms giving a non-zero contribution after
integration in $dz'$ in Eq. (\ref{undurads1}) are those for
$n=-m-3$, $n=-m-1$ and $n=-m-2$. Therefore, we rewrite Eq.
(\ref{undurads1}) as:

\begin{eqnarray}
&&\vec{\widetilde{E}}_{\bot} =\frac{i \omega }{c z_o}
\int_{-\infty}^{\infty} d l'_x \int_{-\infty}^{\infty} d
l'_y\int_{-L_w/2}^{L_w/2} dz'  \exp\left\{i \frac{\omega}{c}
\left[ \frac{z_o\left(\theta_x^2+\theta_y^2\right)}{2} - (\theta_x
l'_x +\theta_y l'_y)\right] \right\} \cr && \times
\widetilde{\rho} \left(z',l'_x,l'_y\right) \exp\left\{i \left[C_2
+\frac{\omega}{2 c} \left(\theta_x-\eta_x^{(c)}\right)^2+
\frac{\omega}{2
c}\left(\theta_y-\eta_y^{(c)}\right)^2\right]z'\right\}\cr
&&\times \sum_{m=-\infty}^{\infty} J_m(u)  \exp\left[-\frac{i m
\pi}{2}\right] \Bigg\{\Bigg[\frac{K}{2i \gamma} \left( J_{-m-3}(v)
\exp\left[-\frac{i 3\pi}{2}\right] - J_{-m-1}(v)
\exp\left[-\frac{i \pi }{2}\right] \right)\cr &&
+\left(\theta_x-\eta^{(c)}_x\right) J_{-m-2}(v) \exp\left[-i \pi
\right]\Bigg]\vec{e}_x+ \Bigg[-\frac{K}{2 \gamma} \left(
J_{-m-3}(v) \exp\left[-\frac{3 i \pi }{2}\right]\right.\cr &&
\left.-
 J_{-m-1}(v) \exp\left[-\frac{i \pi }{2}\right]
\right)+\left(\theta_y-\eta^{(c)}_y\right) J_{-m-2}(v)
\exp\left[-i \pi\right]\Bigg]\vec{e}_y\Bigg\}~, \label{undurads2}
\end{eqnarray}
where we neglected the phase contribution ${K
(\theta_x-\eta_x^{(c)})}/({k_w \gamma})$ because is much smaller
than unity. Non-negligible terms in the expansion of $J_p(\cdot)$
are those for small values of $|p|$, because arguments are much
smaller than unity.

Let us begin with contributions for $m=0$. In this case $J_m(u) =
J_0(u) \sim 1$ and $J_{-m-1}(v) = J_{-1}(v) \sim v$. Thus, terms
in $J_{-m-1}(v)$ scale as $K v/\gamma $. Terms in $J_{-m-3}(v)$
scale as $K v^3/\gamma$ and are negligible. Terms in $J_{-m-2}(v)$
scale as $\left|\theta_{x,y}- \eta_{x,y}^{(c)}\right| v^2 \sim 4 K
v \bar{\gamma}_z \left|\vec{\theta}- \vec{\eta}^{(c)}\right|^2
/\sqrt{1+K^2} \ll K |v|/\gamma$ and are also negligible.

In the case $m=-1$, $J_m(u) = J_{-1}(u) \sim u$ and $J_{-m-1}(v) =
J_{0}(v) \sim 1$. Thus, terms in $J_{-m-1}(v)$ scale as $K
u/\gamma $, that is of the same order of $K v/\gamma $. Terms in
$J_{-m-3}(v)$ scale as $K u v^2/\gamma$ and are negligible. Terms
in $J_{-m-2}(v)$ scale as $\left|\theta_{x,y}-
\eta_{x,y}^{(c)}\right| |u|~ |v| \sim 4 K u  \bar{\gamma}_z
\left|\vec{\theta}- \vec{\eta}^{(c)}\right|^2 /\sqrt{1+K^2} \ll K
|u|/\gamma$ and are also negligible.

Similarly, all other values of $m$ give negligible contributions.
As a result we obtain

\begin{eqnarray}
&&\vec{\widetilde{E}}_{\bot} =\frac{i \omega^2 \left(\vec{e}_x+ i
\vec{e}_y\right)}{2 c z_o \omega_r} \frac{K^2 }{1+K^2}
\left[\left(\theta_x-\eta_x^{(c)}\right) +i
\left(\theta_y-\eta_y^{(c)}\right)\right]\cr && \times
\int_{-\infty}^{\infty}
 d l'_x \int_{-\infty}^{\infty}
d l'_y\int_{-L_w/2}^{L_w/2} dz' \exp\left\{i \frac{\omega}{c}
\left[ \frac{z_o\left(\theta_x^2+\theta_y^2\right)}{2} -
\left(\theta_x l'_x +\theta_y l'_y\right)\right] \right\} \cr &&
\times \widetilde{\rho} \left(z',l'_x,l'_y \right) \exp\left[i C_2
z'\right] \exp\left\{i \frac{\omega}{2c}\left[
\left(\theta_x-\eta_x^{(c)}\right)^2+
\left(\theta_y-\eta_y^{(c)}\right)^2\right]  z'\right\}  ~.
\label{undurads3}
\end{eqnarray}
Note that the electric field is left circularly polarized
(rotating anti-clockwise in the judgement of an observer located
after the undulator and looking towards the device, as it must be)
and vanishes at $\vec{\theta} = \vec{\eta}^{(c)}$. Polarization
characteristics are the same as for the fundamental harmonic,
although the fundamental does not vanish at $\vec{\theta} =
\vec{\eta}^{(c)}$. Spatial resonance is organized along the
undulator for particular values of $m$, as discussed above.

We conclude our analysis of NHG in helical wigglers studying
on-axis harmonic generation. We can do so in all generality, i.e.
for any harmonic number, with the help of Eq. (\ref{undurads1}).
We set $\vec{\theta}-\vec{\eta}^{(c)} =0$, because the concept of
"on-axis emission" is evidently related to the coherent deflection
angle $\vec{\eta}^{(c)}$. We see by inspection that when
$\vec{\theta}-\vec{\eta}^{(c)} =0$, the only non-zero
contributions are for $n=m=0$, because both $u=0$ and $v=0$. Thus,
Eq. (\ref{undurads1}) can be rewritten as

\begin{eqnarray}
\vec{\widetilde{E}}_{\bot} &=& \frac{i \omega }{c z_o} \int
d\vec{l}'  \int_{-L_w/2}^{L_w/2} dz' \widetilde{\rho}
\left(z',\vec{l}' \right) \exp\left[i C_h z'\right] \cr && \times
\left\{\left[\frac{K}{2i \gamma} \Big(\exp[i (h+1) k_w z']-\exp[i
(h-1) k_w z'] \Big) \right]\vec{e}_x\right.\cr &&\left.
+\left[-\frac{K}{2\gamma}\Big(\exp[i (h+1) k_w z']+\exp[i (h-1)
k_w z'] \Big)\right]\vec{e}_y\right\}~. \label{unduradsaxis}
\end{eqnarray}
Since $\widetilde{\rho}$ is a slowly function of $z'$ on the scale
of the undulator period and $C_h \ll k_w$, we see by inspection
that, after integration in $dz'$, one obtains non-zero on-axis
field only for $h=1$, leading to

\begin{eqnarray}
\vec{\widetilde{E}}_{\bot} &=& -\frac{K \omega
(\vec{{x}}+i\vec{e}_y)}{2 c z_o \gamma} \int d\vec{l}'
\int_{-L_w/2}^{L_w/2} dz' \widetilde{\rho} \left(z',\vec{l}'
\right) \exp\left[i C_h z'\right] ~. \label{unduradsaxis2}
\end{eqnarray}
All other harmonics vanish on-axis. This result is in open
contrast with what reported in reference \cite{FREU}. In Section
\ref{sec:criti} we will elaborate further on this issue.

\subsection{\label{sub:mod} Analysis of a simple model}

Let us treat a particular case to exemplify our results. Namely,
let us consider the case when $C_2=0$ (i.e. $\omega = 2 \omega_r$)
and $\widetilde{\rho}\left(z',l'_x, l'_y\right)$ is given by

\begin{equation}
\widetilde{\rho} =\frac{I_o {a}_2}{2\pi c \sigma_\bot^2}
\exp{\left(-\frac{{l'}^{2}_x+{l'}^{2}_y}{2
\sigma_\bot^2}\right)}\exp \left[i \frac{2\omega_{r}}{c}
\left(\eta^{(c)}_x l'_x + \eta^{(c)}_y l'_y \right)
\right]H_{L_w}(z) ~, \label{expara}
\end{equation}
with $H_{L_w}$  a window function equal unity inside the undulator
and zero everywhere else. Here $I_o$ is the bunch current, ${a}_2$
is a constant determining the strength of the bunching and
$\sigma_\bot$ the rms transverse size of the electron beam.

This particular case corresponds to a modulation wavefront
perpendicular to the direction of motion of the
beam\footnote{Information about the modulation wavefront of the
beam is included in $\widetilde{\rho}$.  A phase factor of the
form $\exp{\left[i \omega \alpha ~ \vec{\eta^{(c)}} \cdot
\vec{l}'/c\right]}$ describes a plane wavefront. The case $\alpha
= 1$ that is considered here corresponds to a modulation wavefront
orthogonal to the $z$ axis. When $\alpha =0$ the modulation
wavefront is orthogonal to the direction of propagation. Other
values correspond to a modulation wavefront that is orthogonal
\textit{neither} to the $z$ axis \textit{nor} to the direction of
propagation. This may be the case in SASE XFEL setups with very
long saturation lengths (order of $10^2$ m), where there can be
coherent orbit perturbations. Note that if the beam is prepared in
a different way so that, for instance, the modulation wavefront is
not orthogonal to the direction of propagation of the beam, Eq.
(\ref{undurads3}) retains its validity.}. In this case Eq.
(\ref{undurads3}) can be written as

\begin{eqnarray}
&&\vec{\widetilde{E}}_{\bot } =\frac{ i \omega_r I_o {a}_2
\left(\vec{e}_x+ i \vec{e}_y\right)}{ \pi \sigma_\bot^2 c^2 z_o}
\frac{K^2 }{1+K^2}\left[\left(\theta_x-\eta_x^{(c)}\right) +i
\left(\theta_y-\eta_y^{(c)}\right)\right]\exp{\left[i
\frac{\omega_r}{c} {z}_o({\theta}_x^2+{\theta}_y^2) \right]}\cr
&&\times \int_{-\infty}^{\infty} d l'_x \int_{-\infty}^{\infty} d
l'_y\int_{-\infty}^{\infty} dz' \exp{\left\{-i
\frac{2\omega_{r}}{c} \left[\left({\theta}_x-\eta^{(c)}_x\right)
l'_x + \left(\theta_y-\eta^{(c)}_y\right) l'_y\right]\right\} }
H_{L_w}(z') \cr && \times \exp\left\{i \frac{\omega_r}{c} \left[
\left(\theta_x-\eta_x^{(c)}\right)^2+
\left(\theta_y-\eta_y^{(c)}\right)^2\right] z'\right\}
\exp{\left(-\frac{{l'}^{2}_x+{l'}^{2}_y}{2 \sigma_\bot^2}\right)}
~. \cr &&\label{undurad6}
\end{eqnarray}
Direct calculations yield

\begin{eqnarray}
&&\vec{\widetilde{E}}_{\bot } =\frac{2 i I_o {a}_2 L_w \omega_r
\left(\vec{e}_x+ i \vec{e}_y\right)}{c^2 z_o} \left(\frac{ K^2
}{1+K^2}\right)\left[\left(\theta_x-\eta_x^{(c)}\right) +i
\left(\theta_y-\eta_y^{(c)}\right)\right]\cr &&\times \exp{\left[i
\frac{\omega_r}{c} {z}_o({\theta}_x^2+{\theta}_y^2) \right]}
\exp\left\{-\frac{2 \sigma_\bot^2 \omega_r^2}{c^2}
\left[\left({\theta}_x-\eta^{(c)}_x\right)^2 +
\left(\theta_y-\eta^{(c)}_y\right)^2\right]\right\} \cr && \times
\mathrm{sinc}\left\{\frac{ L_w \omega_r}{2 c}
\left[\left({\theta}_x-\eta^{(c)}_x\right)^2 +
\left(\theta_y-\eta^{(c)}_y\right)^2\right]\right\} ~.
\label{undurad8bis}
\end{eqnarray}
Going back to our particular case in Eq. (\ref{undurad8bis}), a
subject of particular interest is the angular distribution of the
radiation intensity, which will be denoted with $I_{2}$. Upon
introduction of normalized quantities

\begin{eqnarray}
\hat{\theta}_{x,y} &=& \sqrt{\frac{2 \omega_{r}
L_w}{c}}~\theta_{x,y}=\sqrt{8\pi N_w}~ \bar{\gamma}_z
~\theta_{x,y} \cr \hat{\eta}^{(c)}_{x,y} &=&\sqrt{\frac{2
\omega_{r} L_w}{c}}~ \eta^{(c)}_{x,y}= \sqrt{8\pi N_w}
~\bar{\gamma}_z ~\eta^{(c)}_{x,y} 
\label{reduced}
\end{eqnarray}
and of the Fresnel number

\begin{equation}
N= \frac{2 \omega_{r} \sigma_\bot^2}{c L_w} ~,\label{frsnintr}
\end{equation}
one obtains

\begin{eqnarray}
{I}_{2}\left(\left|\vec{\hat{\theta}}-\vec{\hat{\eta}}^{(c)}
\right|\right) &=& \mathrm{const} \times
\left|\vec{\hat{\theta}}-\vec{\hat{\eta}}^{(c)} \right|^2
\exp\left\{- {N} \left|\vec{\hat{\theta}}-\vec{\hat{\eta}}^{(c)}
\right|^2\right\} \mathrm{sinc}^2\left\{\frac{1}{4}
\left|\vec{\hat{\theta}}-\vec{\hat{\eta}}^{(c)}\right|^2\right\}
~. \cr &&\label{I2xybis}
\end{eqnarray}
In the limit for $N \ll 1$, Eq. (\ref{I2xybis}) gives back the
directivity diagram for the second harmonic radiation from a
single particle in agreement with \cite{KIN1}.

\begin{figure}
\begin{center}
\includegraphics*[width=100mm]{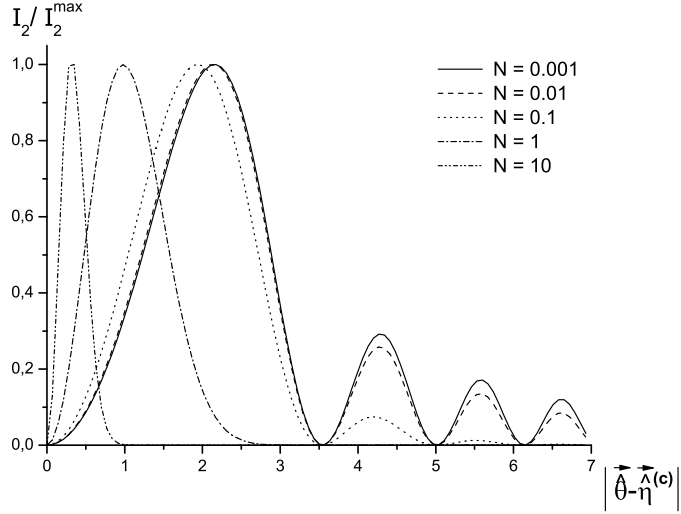}
\caption{Plot of the directivity diagram for the radiation
intensity as a function of $\left|\vec{\hat{\theta}}-
\vec{\hat{\eta}}^{(c)}\right|$ for different values of ${N}$,
normalized to the maximal intensity $I_2^{\mathrm{max}}$ at each
value of $N$. \label{DIRDIA}}
\end{center}
\end{figure}

As an example, the directivity diagram in Eq. (\ref{I2xybis}) is
plotted in Fig. \ref{DIRDIA} for different values of ${N}$ as a
function of $\left|\vec{\hat{\theta}}-
\vec{\hat{\eta}}^{(c)}\right|$, normalized to the maximal
intensity $I_2^{\mathrm{max}}$ at each value of $N$.

The next step is the calculation of the second harmonic power that
is given  by

\begin{eqnarray}
W_{2} &=& \frac{c}{4 \pi} \int_{-\infty}^{\infty} dx_o
\int_{-\infty}^{\infty} dy_o \overline{|\vec{E}_{\bot}(z_o, x_o,
y_o,t)|^2} = \frac{c}{2 \pi} \int_{-\infty}^{\infty} dx_o
\int_{-\infty}^{\infty} dy_o {|\vec{\widetilde{E}}_{\bot}(z_o,
x_o, y_o)|^2}~, \cr &&\label{xpowden}
\end{eqnarray}
where $\overline{(...)}$ denotes averaging over a cycle of
oscillation of the carrier wave and $\vec{E}_{\bot}(z_o, x_o,
y_o,t)$ is the electric field in the space-time domain at position
$(x_o,y_o,z_o)$ and time $t$ .

We will still consider the model specified by Eq. (\ref{expara})
with $C_2=0$. It is convenient to present the expressions for
$W_{2}$ in a dimensionless form. After appropriate normalization
it is a function of one dimensionless parameter only, that is

\begin{equation}
\hat{W}_{2} =    F_2(N) = \ln{\left(1+\frac{1}{ 4
{N}^2}\right)}~.\label{powden2}
\end{equation}
Here $\hat{W}_{2} = W_{2}/W_{o}^{(2)}$ is the normalized power,
while the normalization constant $W_{o}^{(2)}$ is given by

\begin{equation}
{W}_{o}^{(2)}=  \left(\frac{2 K^2}{1+K^2}\right)^2 \frac{I_o^2}{c}
{a}_2^2 ~. \label{W02}
\end{equation}
For practical purposes it is convenient to express Eq. (\ref{W02})
in the form

\begin{equation}
{W}_{o}^{(2)}=  \left(\frac{2 K^2}{1+K^2}\right)^2 W_b {{a}_2^2}
\left(\frac{I_o}{\gamma I_A}\right)~,\label{W02bis}
\end{equation}
where $W_b = m_e c^2 \gamma I_o/e$ is the total power of the
electron beam and $I_A = m_e c^3/e \simeq 17$ kA is the Alfven
current.

\begin{figure}
\begin{center}
\includegraphics*[width=100mm]{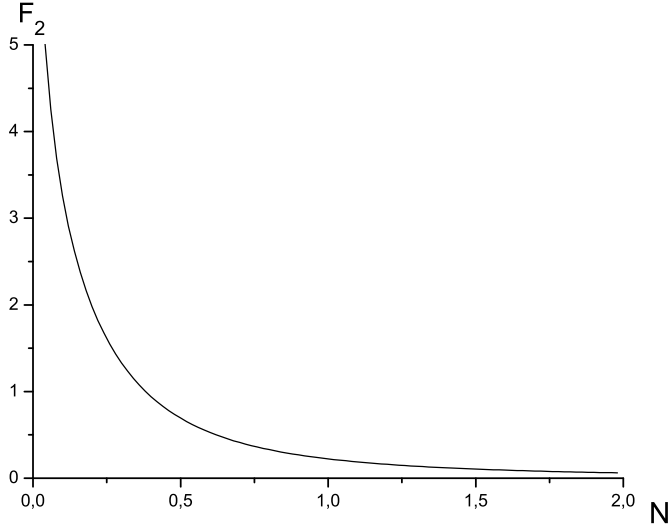}
\caption{Illustration of the behavior of $F_2(N)$. \label{W2}}
\end{center}
\end{figure}

The function $F_2(N)$ is plotted in Fig. \ref{W2}. The logarithmic
divergence in $F_2(N)$ in the limit for $N \ll 1$ imposes a limit
on the meaningful values of $N$. On the one hand, the
characteristic angle $\hat{\theta}_\mathrm{\max}$ associated with
the intensity distribution is given by
$\hat{\theta}_\mathrm{\max}^2 \sim 1/{N}$. On the other hand, the
expansion of the Bessel function performed in Eq.
(\ref{undurads3}) is valid only as $\hat{\theta}^2 \lesssim N_w$.
As a result we find that Eq. (\ref{powden2}) is valid only up to
values of $N$ such that $N \gtrsim N_w^{-1}$. However, in the case
$N < N_w^{-1}$ we deal with a situation where the dimensionless
problem parameter $N$ is smaller than the accuracy of the
resonance approximation $\sim N_w^{-1}$. In this situation our
electrodynamic description does not distinguish anymore between a
beam with finite transverse size and a point-like particle and,
for estimations, we should replace $\ln{(N)}$ with
$\ln{(N_w^{-1})}$.

\section{\label{sec:criti} Discussion}

NHG in a helical wiggler has been addressed in \cite{FREU}. In the
numerical study-case proposed in that reference, an
ultrarelativistic beam at $140$ MeV  is considered, driving an FEL
oscillator operating near  $1~ \mu$m wavelength in free space. NHG
is studied through simulations for an undulator with a uniform
field region of $20$ periods. The complete parameter set can be
found in reference \cite{FREU}. For our purpose, parameters
reported above are enough to guarantee that paraxial and resonance
approximation can be applied to the study-case in \cite{FREU}. In
that reference one can read: "Our conjecture was that on-axis
harmonic excitation due to NHG in helical wigglers can arise
because the nonlinear bunching due to the fundamental creates
on-axis harmonic radiation. This conjecture is borne out in
simulation", and also: "Because NHG is driven by the fundamental,
which excites on-axis modes, we speculate that NHG in helical
wigglers will have substantial on-axis power". In the same
reference one finds a statement regarding harmonic radiation from
a single electron too: "Kincaid \cite{KIN1} showed that
spontaneous generation from helical wigglers vanishes on axis".

Thus, the main result of \cite{FREU} is that characteristics of
helical undulator radiation from an extended source, i.e. a
bunched electron beam, are drastically different compared to those
from a single electron studied in \cite{KIN1}. In the previous
Section we have seen that our conclusions are in open contrast to
\cite{FREU}. In this Section we give reasons why, in our
understanding, results in \cite{FREU} are incorrect.

A first hint follows from general properties of linear
superposition. Any linear superposition of a given field harmonic
from single electrons conserves single-particle characteristics
like parametric dependence on undulator parameters and
polarization. In particular, if a field harmonic from a single
electron vanishes on-axis, it must vanish for the linear
superposition as well. Consider a bunched beam prepared in such a
way that electrons enter the undulator with fixed offsets with
respect to the longitudinal axis and with specific phases related
with their positions along the bunch.  Radiation fields generated
by this beam can be seen as a linear superposition of fields from
individual electrons. Now, as shown in \cite{KIN1}, harmonic
radiation from each of these electrons vanishes on axis.  It
follows that the total field has zero on-axis power as well. This
property is conserved when the dependence of charge and current
density distributions of the sources on the longitudinal
coordinate is slow on the scale of the undulator period. This is
always the case for NHG from ultrarelativistic beams in an FEL
system in free space.

\begin{figure}
\begin{center}
\includegraphics*[width=100mm]{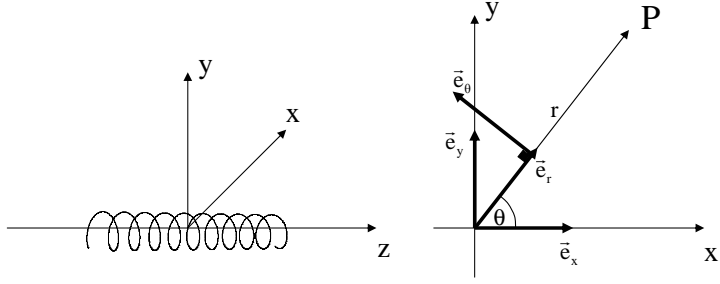}
\caption{Cylindrical coordinate system and undulator
setup.\label{coord}}
\end{center}
\end{figure}

This argument suggests that results in reference \cite{FREU} are
incorrect. A formal demonstration of this fact can be given
discussing the mechanism proposed in \cite{FREU} to explain net
energy exchange between individual electrons and harmonics of the
fundamental in helical undulators. This is obtained through an
azimuthal resonance condition, that is introduced  with these
words\footnote{Note that this mechanism is independent of the type
of harmonic generation process, because it involves interaction
between a single particle and the electromagnetic field. In
particular, it pertains both LHG and NHG.}: "The azimuthal
electron motion in helical wigglers is $\theta = k_w z$ ($k_w$ is
the wave number for the wiggler period $\lambda_w$), which couples
to circularly polarized waves that vary as $\exp(i \phi_h)$, where
$\phi_h = k z + h \theta - \omega t$ is the wave phase. Hence, the
phase along the particle trajectories varies as $\phi_h = (k + h
k_w) z  - \omega t$, and the $h$th order azimuthal mode
corresponds to the $h$th harmonic resonance [i.e., $\omega \approx
(k + h k_w) v_z$]" (cited from \cite{FREU}). Here $\theta$
indicates an azimuthal position as in Fig. \ref{coord}.

It should be remarked that for $h=1$ one obtains $\omega \approx
(k + k_w) v_z$ that is the usual resonance condition between an
azimuthal symmetric wave and an electron in a helical undulator.
This means that the phase $\phi_h = k z + h \theta - \omega t$
pertains a circularly polarized wave whose electric field is
written in terms of unit vectors $\vec{e}_\rho$ and
$\vec{e}_\theta$ in polar coordinates, and not in terms of unit
vectors $\vec{e}_x$ and $\vec{e}_y$ in cartesian coordinates
($\vec{e}_\rho$, $\vec{e}_\theta$, $\vec{e}_x$ and $\vec{e}_y$ are
defined in Fig. \ref{coord}). In fact one may write the electric
field of a (e.g. left) circularly polarized plane wave at position
$\vec{r}$ and time $t$ as :

\begin{eqnarray}
\vec{E}\Big(\vec{r},t\Big) && = E_o\Big(\vec{e}_x + i~ \vec{e}_y
\Big) \exp\Big[i k z + i (h-1) \theta - i \omega t\Big] \cr && =
E_o\Big(\vec{e}_r + i ~\vec{e}_\theta \Big) \exp\Big[i k z + i h
\theta - i \omega t\Big]\equiv E_o\Big(\vec{e}_r + i
~\vec{e}_\theta \Big) \exp\Big[i \phi_h\Big]~,\label{thetaer}
\end{eqnarray}
$E_o$ being a constant field strength. This definition of $\phi_h$
is in agreement with notation in \cite{FREB,FREO}, and justify
words in \cite{FREU}: "the $h$th order azimuthal mode corresponds
to the $h$th harmonic resonance". In fact, we can write the angle
$\psi$ between the transverse velocity of the electron and the
transverse electric field vector as  $\psi = k_w z + [k z+ (h-1)
\theta - \omega t]$. Authors of \cite{FREU,FREB} use relation
$\theta \approx k_w z $ in the expression for $\phi_h$, and
automatically in the expression for $\psi$ too. Using also $dz =
v_z dt$, where $v_z\simeq c$ is constant for an electron in a
helical undulator, they obtain resonance for $d\psi/(d z) = k + h
k_w - \omega/v_z = 0$. As said before, for the particular case
$h=1$ (fundamental harmonic) one has $h-1=0$ (azimuthal-symmetric
wave) and $d\psi/(d z)=0$ for $\omega = (k + k_w) v_z$, that is
the usual resonance condition.

We argue that the azimuthal resonance described above is a
misconception arising from a kinematical mistake. Namely, it is
incorrect to use relation $\theta \approx k_w z$ in the expression
for $\psi$ as done in \cite{FREU,FREB}. Let us show this fact.

\begin{figure}
\begin{center}
\includegraphics*[width=100mm]{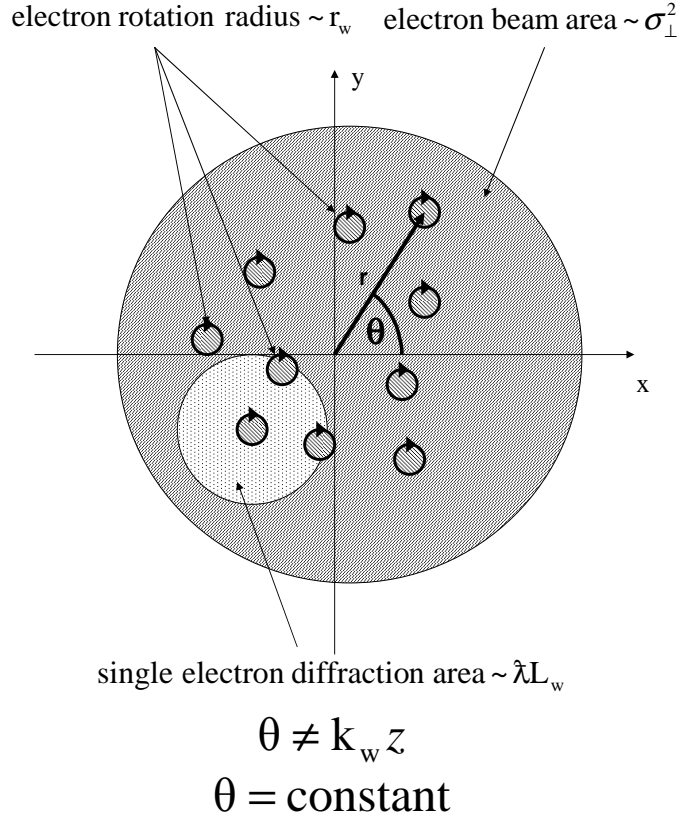}
\caption{Hierarchy of characteristic scales of interest.
\label{coor2}}
\end{center}
\end{figure}

We begin introducing some characteristic scale of interest. On the
one hand, the radiation diffraction size for a single particle is
of order $\sqrt{\lambdabar L_w}$. On the other hand, the electron
rotation radius is given by $r_w = (K/\gamma) \lambdabar_w$. It
follows that

\begin{eqnarray}
\frac{r_w^2}{\lambdabar L_w} = \frac{1}{\lambdabar
L_w}\frac{K^2}{\gamma^2}\lambdabar_w^2 = \frac{1}{\lambdabar
L_w}\frac{2 K^2}{1+K^2} \lambdabar \lambdabar_w = \frac{K^2}{\pi
N_w (1+K^2)} \ll 1~,  \label{scale}
\end{eqnarray}
where we used the fact that $\gamma^2 = (1+K^2) \bar{\gamma}_z^2$,
where  $\bar{\gamma}_z$ has been introduced in Eq.
(\ref{gamzdef}). Inequality (\ref{scale}) holds independently of
the value of $K$, because $N_w \gg 1$. Thus, the electron rotation
radius is always much smaller than the radiation diffraction size.
To complete the picture we introduce a last characteristic scale
of interest in our problem, pertaining the electron beam rather
than a single electron. This scale is the transverse size of the
electron beam, $\sigma_\bot$. Straightforward geometrical
considerations show that it makes sense to talk about
$\sigma_\bot$ only when $\sigma_\bot^2 \gg r_w^2$. This is the
case in practical situations of interest. There is still room to
compare the beam size $\sigma_\bot$ with the radiation diffraction
size. In the case $\sigma_\bot \ll \sqrt{\lambdabar L_w}$ we deal
with a filament electron beam, and results in \cite{KIN1} can be
directly applied. In the opposite case the filament beam
approximation breaks down. We are interested in this last case,
where single-particle results cannot be used. We have therefore
established a hierarchy in the characteristic scales of interest:
$\sigma_\bot^2 \gtrsim \lambdabar L_w \gg r_w^2$. The situation is
described in Fig. \ref{coor2}, where we schematically indicated
the electron beam area and the diffraction area as disks. In Fig.
\ref{coor2} we also indicate the azimuthal and radial coordinates
of a single electron, respectively $\theta$ and~$r$. As one can
see, the azimuthal coordinate of each electron, $\theta$, is fixed
during the motion inside the undulator with the accuracy of the
resonance approximation, scaling as $1/N_w$.  In contrast to this,
the identification $\theta \approx k_w z$ is made in
\cite{FREU,FREB}. This is a kinematical mistake. If $\theta
\approx k_w z$ each electron would be rotating around the origin
of the coordinate system, that is not the case. Thus, we have
shown that $\psi = k_w z + [k z+ (h-1) \theta - \omega t]$  with
$\theta$ constant for each electron along the trajectory in the
undulator. As a result $d\psi/(dz) = 0$ only for $h=1$ that yields
the usual resonance condition $\omega = (k + k_w) v_z$. Summing
up, our conclusion is that a kinematical mistake is at the origin
of the azimuthal resonance described in \cite{FREU,FREB}. The
azimuthal resonance condition is a misconception following from
this mistake. This misconception is subsequently passed on to
simulations in \cite{FREU}, resulting in incorrect outcomes.
Harmonic emission exists for a single electron \cite{KIN1}, and it
also exists for an electron bunch. However, qualitative properties
are different with respect to what has been predicted in
\cite{FREU}. In particular, as we have seen in Section
\ref{sec:theor}, on-axis power vanishes.

\section{\label{sec:conc} Conclusions}

In this paper we discussed NHG in helical undulators, with
particular emphasis on second harmonic generation. First we
discussed the NHG mechanism in helical undulators in all
generality. Then we specialized our study to the case of second
harmonic generation. Finally, to exemplify our results, we treated
a simplified model where the beam modulation wavefront is
orthogonal to the $z$ axis, has a Gaussian  transverse profile and
is independent on the position inside the undulator.

Our results show that on-axis harmonic generation from helical
wigglers vanishes. This applies to any harmonic of interest with
the exclusion of the fundamental, and independently of the form of
the electron beam modulation (assuming that the electron beam as a
whole propagates on-axis). Important consequences follow regarding
the two mainstream development paths in FEL physics. First, as
concerns short wavelength (x-ray) SASE FEL devices, vanishing
on-axis harmonics make the option of a helical undulator less
attractive as regards the exploitation of NHG radiation. Second,
as concerns high average-power FEL oscillators, vanishing on-axis
harmonics suggest that helical undulators carry relevant
advantages over planar undulators, as potential for mirror damage
is reduced.

Previous studies reported non-vanishing on-axis power, due to the
nature of a particular azimuthal resonance condition. We showed
that this resonance condition is a misconception, arising from a
kinematical mistake. This misconception was passed on to
simulations, that confirmed the presence of on-axis power out of
NHG from helical wigglers. This result is also incorrect.
Simulations are undoubtedly very important scientific tools, but
they follow specific models. The correctness of their outcomes is
related to the correctness of these models,  meaning that validity
of simulations should always be cross-checked with analytical
results from simplified study-cases.

\section{\label{sec:ackn} Acknowledgements}

We thank Martin Dohlus (DESY) for useful discussions, Massimo
Altarelli (DESY) and Jochen Schneider (DESY) for their interest in
this work.

\end{document}